\documentclass[12pt]{article}
\usepackage{fullpage,amssymb}

\newcommand\pr{\prime}
\newcommand\be{\begin{equation}}
\newcommand\ee{\end{equation}}
\newcommand\bea{\begin{eqnarray}}
\newcommand\eea{\end{eqnarray}}
\newcommand\bean{\begin{eqnarray*}}
\newcommand\eean{\end{eqnarray*}}

\newcommand\bdm{\begin{displaymath}}
\newcommand\edm{\end{displaymath}}

\def\pmb#1{\setbox0=\hbox{#1}%
  \kern-.025em\copy0\kern-\wd0
   \kern.05em\copy0\kern-\wd0
   \kern-.025em\raise.0433em\box0 }

\def\rd{\mathrm d}

\begin{document}

\title{\large {\bf  The generalized F\'enyes--Nelson model for free scalar field theory\thanks{{\em Letters in Mathematical Physics\/} {\bf 4}  (1980) 101--106. \copyright  1980 {\em D.\ Reidel Publishing Company.}}}}

\author{{\normalsize Mark Davidson}\thanks{
Current Address: Spectel Research Corporation, 807 Rorke Way, Palo Alto, CA   94303 
\newline  Email:  mdavid@spectelresearch.com, Web: www.spectelresearch.com}\\
\normalsize{\em Department of Physics, San Jose State University, San Jose, CA  95192, U.\ S.\ A.}}

\date{}

\maketitle
\begin{abstract} 
The generalized F\'enyes--Nelson model of quantum mechanics is applied to the free scalar field.  The resulting Markov field is equivalent to the Euclidean Markov field with the times scaled by a common factor which depends on the diffusion parameter.  This result is consistent between Guerra's earlier work on stochastic quantization of scalar fields.  It suggests a deep connection between Euclidean field theory and the stochastic interpretation of quantum mechanics.  The question of Lorentz covariance is also discussed.
\end{abstract}

\section{Introduction}

In this paper a class of stochastic models for the quantized free scalar field is presented.  These models are direct descendents of the Markov model for quantum mechanics proposed by Imre F\'enyes \cite{fenyes1} almost three decades ago.  Important and elegant contributions to this subject were made by Nelson \cite{nelson1,nelson2} who greatly expanded on the work of  F\'enyes and developed a powerful mathematical framework for the study of stochastic models of quantum mechanics.  Nelson also made great contributions to Euclidean and constructive field theory \cite{nelson3}--\cite{simon1}, and in this paper it is shown that Euclidean field theory and stochastic models of quantum mechanics are intimately related.  The first statement of this result was made by Guerra \cite{guerra1} who showed that Euclidean field theory results from the stochastic quantization of the free scalar field.

A generalization of Guerra's result is presented here, and the question of Lorentz covariance is investigated.  The approach taken is to apply the  generalized F\'enyes--Nelson model \cite{davidson1} (here-after GFN model) to the free scalar field.  In this model the diffusion parameter is undetermined.  For a special value of the diffusion parameter Guerra's result is recovered.  If one continues the diffusion parameter to imaginary values, as in \cite{davidson2}, then the full mathematics of quantum field theory are recovered.  The justification for this continuation is that it is believed that measurable quantities must be independent of the diffusion parameter.  This conclusion has been strongly supported by Shucker's work on the free particle Schr\"odinger equation \cite{shucker1}.  It is believed that measurable quantities will be independent of the diffusion parameter for more general dynamical systems including fields, although a detailed examination of this question has not yet appeared in print.

Upon applying the GFN model to the free scalar field one finds that the expectations of the Markov theory are equal to the Schwinger functions for free scalar theory with all times in the Schwinger functions scaled by a common factor.  When the diffusion parameter is continued to imaginary values, the Markov expectations become the causal Green's functions of quantum field theory which are Lorentz invariants.  The conclusion is that if the diffusion parameter is indeed not measurable, then the stochastic model for the field is consistent with special relativity, although it is not manifestly Lorentz covariant.

\section{The free scalar field}

Consider the classical equation for a real free scalar field in Minkowski space:
\be
P_\mu P^\mu\phi =m^2\phi, \qquad P_\mu =-i\hbar \frac{\partial}{\partial x^\mu},
\label{1}
\ee
where $c$ has been set to unit and $g^\infty =+1$.  Demand periodic boundary conditions on the field:
\be
\phi(x, t) =\phi (x +a, t), \qquad a_i =n_i L, \qquad n_i  \mbox{ integer}
\label{2}
\ee
so that one may write:
\be
\phi (x, t) =\left(\hbar L^3\right)^{-1/2}\sum\limits_k \exp \left( i\vec{k} \cdot \vec{x}\right) \phi_k (t), \qquad k_i =2\pi m_i /L, \qquad m_i \mbox{ integer}
\label{3}
\ee
where the factor preceding the sum has been included as a convenient normalization factor.  Reality demands:
\be
\phi_k (t) =\phi^*_{-k} (t).
\label{4}
\ee
The equations of motion for the field components are:
\be
\left[\hbar^2 \frac{\partial^2}{\partial t^2} +\left(\hbar^2 \vec{k}^2 +m^2\right)\right] \phi_k (t)= 0.
\label{5}
\ee
Let us    write:
\be
\phi_k (t) =R_k (t) +iI_k (t), \qquad R \mbox{ and } I \mbox{ real }
\label{6}
\ee
then reality demands
\be
R_k (t) =R_{-k} (t),\qquad I_k (t) =-I_{-k} (t).
\label{7}
\ee

The GFN model \cite{davidson1} shall now be applied to this system.  Only the ground state of the field will be considered.  In the ground state, all components of the field are independent, except for the dependency required by (\ref{7}).  $R$ and $I$ become random variables satisfying the stochastic differential equations:
\bea
\rd R_k (t) &=& b^1_k (R_k) \rd t + \rd W^1_k (t),\label{8}\\
\rd I_k (t) &=& b^2_k (I_k) \rd t + \rd W^2_k (t),
\label{9}
\eea
where the conditions in eqn.\ (\ref{7}) require
\be
b^1_k =b^1_{-k}, \qquad b^2_k = -b^2_{-k}
\label{10}
\ee
and 
\be
W^1_k =W^1_{-k}, \qquad W^2_k =-W^2_{-k}.
\label{11}
\ee
The $W$'s in eqns.\ (\ref{8}) and (\ref{9}) are Wiener processes satisfying
\bea
E(\rd W^1_k (t) \rd W^1_{k^\pr} (t)) &=& 2\nu \rd t (\delta_{k,k^\pr} + \delta_{k,-k^\pr})\label{12}\\
E(\rd W^2_k (t) \rd W^2_{k^\pr} (t)) &=& 2\nu \rd t (\delta_{k,k^\pr} -\delta_{k, -k^\pr})
\label{13}
\eea
and where $W^1$ and $W^2$ are independent of one another.

The ground state probability densities for the $R_k$ and $I_k$ are derived from  Schr\"odinger's equation.  They are found to be (up to a normalization constant)
\be
\rho_k (x) =\exp \left[ -\left( 2E_k/\hbar^2\right) x^2 \right] ,\qquad E_k =\frac{1}{2} \hbar \sqrt{\vec{k}^2 +m^2 /\hbar^2}
\label{14}
\ee
where $x$ can be either $R_k$ or $I_k$.

Following the prescription presented in Ref.\ 8 for constructing a stochastic process, one writes:
\be
b^1_k =\nu \frac{\partial}{\partial R_k} \ln (\rho_k (R_k)), \qquad b^2_k =\nu \frac{\partial}{\partial I_k} \ln (\rho_k (I_k))
\label{15}
\ee
where $\nu$ can take on any value from $0$ to $\infty$.  One finds:
\be
b^1_k =-4\nu (E_k/\hbar^2) R_k, \qquad b^2_k =-4\nu (E_k /\hbar^2) I_k.
\label{16}
\ee
These expressions for $b^1$ and $b^2$ satisfy a global Libschitz condition so that existence and uniqueness solutions  to (\ref{8}) and (\ref{9}) are ensured by the Piccard method \cite[p.\ 43]{nelson1}.  In can be easily shown that both $I_k$ and $R_k$ are Gaussian Markov processes, so that all expectations of products are generated by the expectations:
\be
E(R_k(t) R_k(t^\pr))=E(I_k (t) I_k (t^\pr)) =\exp \left[ -\lambda | t-t^\pr|\right] \hbar^2 /4 E_k, \qquad \lambda =4\nu E_k /\hbar^2,
\label{17}
\ee
and where these results may be easily derived by integrating (\ref{8}) and (\ref{9}), multiplying and taking an expectation.

We now come to the main point.  Recall the Schwinger functions for free scalar fields
\bea
S_2(x_1, t_1 ; x_2, t_2) &=& \int \frac{\rd^4 k}{(2\pi)^4} \, 
\frac{ \exp \left[i\vec{k} \cdot \left( \vec{x}_1 -\vec{x}_2\right) +ik_0 \left(t_1 -t_2\right)\right]}{\left(k^2_0 +\vec{k}^2 +m^2/\hbar^2\right)},\label{18}\\
S_N\left(x_1, t_1;\ldots; x_N, t_N\right) &=& \sum\limits_\pi S_2 \left( x_1, t_1 ; x_2, t_2\right)\times \ldots \times S_2 \left( x_{N-1}, t_{N-1}; x_N , t_N\right),
\label{19}
\eea
where the sum over $\pi$ is a sum over distinct permutations of the arguments of the $S$'s.  We state our result as a theorem, but we leave it to the reader to prove (\ref{17}).

\noindent {\sc Theorem}. {\em For non-coincident $x_i$:
\be
\lim\limits_{L\to \infty\atop  \Lambda\to \infty} E\left(\phi\left(x_1, t_1\right) \times \ldots\times \phi \left(x_N, t_N\right)\right) =S_N \left( x_1, \frac{2\nu}{\hbar}\, t_1 ; \ldots ; x_N , \frac{2\nu}{\hbar}\, t_N\right)
\label{20}
\ee
where $\Lambda$ is a cutoff in $\vec{k}^2$.
}

{\em Proof}.  To show this result, it suffices to prove (\ref{20}) for the two point function only because the expectations of (\ref{20}) will satisfy    eqn.\ (\ref{19}) since $\phi$ is a Gaussian process.  Using eqn.\ (\ref{3}) we have 
\be
E\left( \phi \left( x_1, t_1 \right) \phi \left( x_2 , t_2\right)\right) =\frac{1}{\hbar L^3} \; \sum\limits^\Lambda_{k,k^\pr} \exp \left( i\vec{k}\cdot \vec{x}_1 +i\vec{k}^\pr \cdot \vec{x}_2 \right) E \left(\phi_k\left( t_1\right)\phi_{k^\pr} \left(t_2\right)\right)
\label{21}
\ee
where a cutoff in $k$ space has been included in eqn.\ (\ref{21}).  From eqn.\ (\ref{17}) it follows
\be
E\left( \phi_k\left( t_1\right) \phi_{k^\pr} \left( t_2\right)\right) =\delta_{k,-k^\pr} 2\exp \left[-\lambda \big|t_1-t_2\big|\right]\hbar^2 /4E_k.
\label{22}
\ee
Substitution into (\ref{21}) then yields
\bdm
E\left( \phi\left( x_1, t\right) \phi \left( x_2, t\right)\right) =\frac{1}{L^3} \sum\limits^\Lambda_k \; \exp \left[i\vec{k} \cdot \left( \vec{x}_1 - \vec{x}_2\right)\right] \, \int \, \frac{\rd k_0}{2\pi} \; \frac{\exp \left[ ik_0 |t_1-t_2 | (2\nu /\hbar )\right]}{\left( k^2_0 +\vec{k}^2 +m^2/\hbar^2\right)}.
\edm
Taking the limits $L\to \infty$ and $\Lambda\to \infty$, this becomes
\bdm
\lim\limits_{L\to \infty\atop \Lambda\to \infty} E(\phi (x_1, t_1) \phi (x_2, t_2)) = S_2 \left( x_1, \frac{2\nu}{\hbar}\, t_1 ; x_2, \frac{2\nu}{\hbar}\, t_2\right)
\edm
and the result is shown.\hfill$\blacksquare$

Note that if $\nu =\hbar /2$, which was the value used by Guerra \cite{guerra1}, then the expectations of the Markov theory become equal to the Schwinger functions which was exactly Guerra's result.

Our theorem shows that the expectations of the generalized F\'enyes--Nelson model have the same analyticity properties in $\nu$ as the Euclidean field's Schwinger function does in $t$.  It is believed that $\nu$ is not an observable and that measurable quantities are independent of $\nu$.  Therefore, one may consider continuing $\nu$ into the complex plane, as was done in \cite{davidson2}.

To illustrate the compatibility of the present theory with special relativity, we continue to the point
\be
\nu =i\hbar /2.
\label{23}
\ee
We denote the analytically continued expectations by $E_\nu$.  One finds easily:
\be
E_{\nu =i\hbar /2} (\phi (x, t) \phi(0)) =i\int \,\frac{\rd^4 k}{(2\pi)^4} \, \frac{\exp (ik^\mu x_\mu)}{\left[ k^\mu k_\mu -\left( m^2 /\hbar^2\right)+i\epsilon\right]}
\label{24}
\ee
where the convergence factor $\epsilon$ is taken to zero after the integral is evaluated using contour integration.  Now compare (\ref{24}) with the causal Green's function of ordinary quantum theory (see Ref.\ 11, p.\ 142, or Ref.\ 12, Appendix C):
\be
D_C (x) =i \langle 0|T\phi (x) \phi(0) |0\rangle =\int \,\frac{\rd^4k}{(2\pi)^4} \; \frac{\exp(ik^\mu x_\mu)}{m^2 -k^\mu k_\mu -i\epsilon}.
\label{25}
\ee
The two theories are related by
\bea
E_{\nu=i\hbar /2} (\phi (x,t)\phi (0)) &=& -iD_C (x) = \langle 0|T\phi (x) \phi (0) |0\rangle,\label{26}\\
E_{\nu =-i\hbar /2} (\phi(x,t) \phi (0)) &=& iD^*_C (x) = \langle 0 |T^* \phi (x) \phi (0) |0\rangle,
\label{27}
\eea
where $T$ and $T^*$ denote time ordering (later times to the left) and anti-time ordering, respectively.  

Using the combinatorial rule of eqn.\ (19) which is satisfied by quantum expectations as well as the Markov expectations and Schwinger functions, we obtain immediately a generalization of (\ref{26}) and (\ref{27}) to  higher-order Green's functions:
\bea
E_{\nu=i\hbar /2} \left(\phi(x_1\right)\times \ldots \times \phi(x_N)) &=&\langle 0|T\phi (x_1) \times \ldots \times \phi (x_N) |0\rangle,\label{28}\\
E_{\nu=-i\hbar /2} (\phi (x_1) \times \ldots \times \phi (x_N)) &=& \langle 0|T^* \phi (x_1) \times \ldots \times \phi (x_N)|0\rangle, 
\label{29}
\eea
where we have stopped writing the $t$ arguments explicitly on the left-hand sides.  Since the right-hand sides of eqns.\ (\ref{28}) and (\ref{29}) are Lorentz invariants it follows that the GFN model for the free scalar field is manifestly Lorentz covariant when continued in the diffusion parameter to $\pm i\hbar /2$.

It should be realized that the above result is a special case of a more general statement which applies as well to non-steady state situations.  When $\nu$ is continued to $\pm i\hbar /2$ in a GFN model, whether steady state or not, then all of the mathematics of ordinary quantum mechanics are recovered as was shown in Ref. 9 for the single particle  Schr\"odinger equation.

\section{Concluding remarks}

The GFN model for a free scalar field is a Markov field whose correlations are obtained from the Schwinger functions by   scaling the times by a factor of $2\nu /\hbar$.

Exploiting the assumed indeterminate nature of $\nu$ leads to the conclusion that the GFN   stochastic model is consistent with the special theory of relativity although it is not manifestly Lorentz covariant except at nonphysical values of the diffusion constant.

I am strongly of the opinion that these  same results are true for the ground state of the interacting theory as well, perhaps with regularity conditions on the form of the interaction.

An interpretation is possible of the continuation from real to imaginary times which has proved so useful in quantum field theory.  This continuation can be thought of as simply   exploiting the indeterminate nature of the diffusion parameter in the GFN stochastic model.

\bigskip

\noindent ({\em Received  January 3, 1980})

\end{document}